\documentclass[sigconf]{acmart}
\usepackage{booktabs}
\usepackage{multirow}
\usepackage{array}
\usepackage{threeparttable}
\usepackage{dblfloatfix}

\AtBeginDocument{%
  }

\setcopyright{acmlicensed}
\copyrightyear{2026}
\acmYear{2026}
\acmDOI{XXXXXXX.XXXXXXX}

\acmISBN{978-1-4503-XXXX-X/2018/06}

\title{Mind Perception Influences Perceived AI Companionability}
\renewcommand{\shortauthors}{Banks and Li}

\author{Jaime Banks}
\authornote{Both authors contributed equally to this research.}
\email{banks@syr.edu}
\orcid{0000-0002-7598-4337}
\affiliation{%
\department{School of Information Studies}
  \institution{Syracuse University}
  \city{Syracuse}
  \state{New York}
  \country{USA}
}

\author{Zhixin Li}
\authornotemark[1]
\email{zli457@syr.edu}
\orcid{0009-0006-1118-4361}
\affiliation{%
\department{School of Information Studies}
  \institution{Syracuse University}
  \city{Syracuse}
  \state{New York}
  \country{USA}
}

\begin{abstract}
Humans increasingly keep company with AI companions, yet whether mind perception (MP) precedes machine companionship remains untested. In a field-sampled experiment, participants received mind-attributive or mind-negating descriptions of an AIC before interacting with and rating its companionability. Baseline skepticism was high; minded primes somewhat reduced skepticism for connective coordination (CC; dyadic co-presence) but not eudaimonic exchange (EE; self-elevating links). Pre-interaction MP beliefs predicted companionability dimensionally: Agentic MP predicted EE potential while experiential MP predicted CC potential. Meaning in life moderated the agentic MP-companionability link, but exclusively for those already high in existential purpose. Results suggest two construal pathways to companionability—one through perceiving the AIC as an agent-resource and one through perceiving it as an attuning experiencer—both more accessible to humans already flourishing.
\end{abstract}

\keywords{Mind perception, conversational agents, social cognition, wellbeing, anthropomorphism, priming}

\setcopyright{none}                             
\renewcommand\footnotetextcopyrightpermission[1]{} 

\begin{document}
\maketitle

\pagestyle{fancy}
\fancyhf{}                                   
\fancyhead[L]{\small Mind Perception Influences Perceived AI Companionability}
\fancyhead[R]{\small Banks and Li}
\fancyfoot[C]{\thepage}                      
\renewcommand{\headrulewidth}{0pt}          

\section{Introduction}
Since the mainstreaming of large language models (LLMs), news headlines are thick with tragedies and lawsuits in which humans interact with AI companions (AICs) to some unfortunate end (e.g., \cite{montoya2026}). In parallel, humans in relationships with AICs defend their value \cite{novozhilova2026}, while others insist interacting with social AI is “sharing minds with the mindless” (\cite{saetra2020}, p. 322)—a concern that some extend to today’s AICs—recognizing them as not genuinely sentient \cite{schwitzgebel2023}. Ontological debates about LLMs (and the technologies they support) have positioned “model welfare” and potential sentience against the notion that machine-mind discourses are a cognitive “trap” that deflects attention from AI dangers \cite{choudhury2026} and commercial interests. Meanwhile, prominent AI scientists and companions are in public disagreement about whether anything akin to consciousness is even achievable in current or future technologies \cite{chalmers2024}. Across these tensions is an important but overlooked operation: Regardless of whether or not AICs do have minds, the human perception that they do underlies both their relational potentials and problematics. 

AICs are multimodal, LLM-driven applications with affordances for customizable persona-building, memory-supported connections over time, conveying emotionally and sexually intimate exchanges, and sustained feelings of connectedness. Dedicated AIC apps, as well as general-purpose LLMs, are increasingly engaged for companionship—for keeping company, intellectual and emotional intimacy, and romantic and sexual relationships. Humans with AICs report sustained and meaningful emotional connections \cite{skjuve2023} and some evidence suggests engaging AICs can offer at least temporary reductions in loneliness comparable to interacting with humans \cite{defreitas2026}. That humans can meaningfully connect with AI comes alongside other evidence that they implicitly (and sometimes explicitly) attribute mental capacities or minded states to AI—like thinking, feeling, planning \cite{colombatto2025}—even while acknowledging they are made objects \cite{manoli2026}. 

These two literatures—mind perception (MP) and machine companionship (MC)—are often discussed in tandem, but the relationship between the phenomena is rarely explicitly tested. A scoping review of machine companionship literature documented ascribed humanlike, minded agency as among the most commonly cited theoretical frameworks, yet empirical examination of MP within MC is largely limited to considerations of functional agency and autonomy \cite{banks2026}. In this way, MP appears to function as a background assumption or justification for why human-machine relations are possible or gratifying. Whether construing AI as a minded entity actually sets the social-cognitive conditions for machine companionship is an open empirical question. Must a human see an AI as a minded “someone home” in order to construe it as companionable?

As an early step toward bridging that gap, we conducted a two-condition experiment in which non-expert participants were primed with a minded or mindless description of an AI companion before interacting with it for 15 minutes and then assessing its companionability. We observe shifts in perceived AIC companionability based on minded or mindless primes and based on pre-interaction beliefs about AI’s mental capacities, with wellbeing conditionally moderating those shifts. We interpret findings to suggest there are two pathways by which social-cognitive construals (both from primes and from a priori beliefs) may give rise to perceived companionability: One through perceiving the AIC as an agent-resource and one through perceiving it as an attuning experiencer. These MP-companionability links may only function for humans that are already flourishing, with implications for AIC design justice.    

\section{Theoretical Background}
\subsection{Mind Perception for Machines}

Mind perception is the construal of an entity as having mental capacities, such as the ability to think, desire, feel, or suffer. Malle \cite{malle2019} offered evidence for a three-factor model for MP, comprising perceived affective, social/moral, and reality-interaction capacities. Weisman and colleagues \cite{weisman2017} also argued for three factors, but related to body, heart, and mind model. We here adopt a Gray and colleagues’ \cite{gray2007} foundational model based on evidence of two MP dimensions: Perceptions of agency (capacities to do, like communicate, remember, plan, judge, and control behavior) and of experience (capacities to feel, like pain, pleasure, joy, fear, and hunger). They demonstrated that MP differs across types of entities (e.g., low experience and moderate agency for a robot, high experience but low agency for a dog, and both high for a typical human) and note the dimensions’ importance to how people think about others in social interactions. For instance, seeing entities as agentically minded is linked to judgments of moral responsibility, while seeing them as experientially minded is linked to perceived capacities to be harmed by others’ behavior \cite{gray2012}. 

Human construals of and relationships with AICs are complex, forming at the intersection of beliefs about AIC agency and identity, humans’ own behaviors, the AI’s behaviors, platform and model dynamics, and social norms and knowledge \cite{lee2026}. Central to this complexity is whether people construe machines as having minds at all. It is well-evidenced that machine MP varies by who is judging, which machine is being judged, and the moment of judgment—the same person may perceive mind in an AI in one context and deny it in another. At one pole, humans see LLMs’ emotional capacity as on par with that of a rock \cite{ladak2025}. At the other, a majority of humans may see AI as having authentic mental states or even consciousness \cite{colombatto2024}. Construals of AI mind (including higher-order attributions of consciousness or sentience) can be animated by AICs’ relationship-seeking behavior, emotional dialog, and relational memory \cite{krik2026} and by apparent developmental cycles mimicking life trajectories \cite{akiyoshi2024}. They also correspond with a host of psychographic factors (e.g., decreasing with adult age, decreasing with need for cognition; \cite{li2026}. MP for AI may impact perceptions of moral responsibility, both positive and negative emotions, satisfaction with interactions, and the formation of social connections \cite{li2026}. It may facilitate harm through problematic behaviors, from verbal abuse and misinformation to normalizing self-harm or risky behaviors \cite{zhang2025}. Humans default to perceiving mental capacities in machines (especially under high cognitive load) as a shortcut for explaining their behaviors \cite{spatola2022}. Social AI may itself move people to engage philosophical and ontological questions of minds and souls and consciousness and how they are linked to one’s purpose and meaning in life \cite{banks2025}. 

Relevant to this investigation, MP for artificial agents is malleable—it can be brought to mind by cues suggesting a machine has motivations (influencing character evaluations; \cite{pataranutaporn2023}), shaped by media representations (correspondent with collaborating and trust; \cite{banks2020}), and prompted by simple descriptions of AI as autonomous or sentient \cite{pauketat2026}. Further, divergence between implicit, behavioral indicators of MP and explicit, verbal indicators of MP (e.g., \cite{keijsers2022}) suggests we may see minds in machines while situated in interactions, even if we would deny holding that belief outside of the interaction context. Alternatively, the implicit and explicit construals may split the MP dimensions, as when most implicit indicators of embodied-AI mind perception (e.g., classic Sally-Anne and behavior prediction) split from explicit mind ascription \cite{thellman2022}, with only implicit affective-mind perceptions corresponding with the explicit attribution \cite{koban2024}. Other evidence links social treatment of machines to implicit, automatic forms of humanization but not to explicit humanization \cite{zlotowski2018}. 

\subsection{Machine Companionship}

Companionship, generally, has been characterized as an ongoing process of reliable, pleasurable exchange \cite{rook1987}, and the notion that an entity as “companionable” can be understood as the designed or perceived suitability of an entity for the purpose of providing persistent, pleasurable exchange. Machine companionship, specifically, is an “autotelic, coordinated connection between a human and machine that unfolds over time and is subjectively positive” \cite[p.~13]{banks2026}. Human-machine relations roughly follow trajectories common to other relationships, including relationship formation through progressive self-disclosure, emotional exploration \cite{skjuve2021}, humanlikeness, and authenticity \cite{pentina2023} and relationship deterioration as a function of dwindling social processes (e.g., attraction, self-disclosure, intimacy; \cite{croes2021}). The speed of that development and deterioration can match or exceed that for human relations (see \cite{kleinert2026}). Machine companionship itself is distinct from what an individual gets from it, such as the satisfaction of one’s emotional needs \cite{rook1987} and non-trivial wellbeing outcomes \cite{defreitas2026}.

An emerging model for MC \cite{banks2026c} suggests the phenomenon has two dimensions. The first, eudaimonic exchange (EE), comprises the meaningfulness of the collaborative connection, fostering feelings that of being inspired, fulfilled, and bettered. The second, connective coordination (CC), involves operational bonding characterized by feeling synchronized with the machine in co-present, co-active care and experience. That model suggests there are divergent templates for companionship—one autotelic and one socioinstrumental—such that companionship with AI may be motivated differently across relationships.  

\subsection{Links Between Mind Perception and Companionship}

It has long been argued that humans, ourselves, mindlessly engage social scripts in our interactions with machines \cite{nass2000} without necessarily attributing mind and even while denying machines are minded \cite{koban2023}. Through that lens, MP may not be necessary for social relations with AI, since relationship-relevant interactions are driven by scripts rather than mediated by nuanced social-cognitive processes. On the other hand, the same perspective suggests it is the machine’s social cues that trigger humans’ script-driven reactions \cite{nass1999}. That is, a human sees a machine doing a social thing (e.g., expressing polite words, engaging in helping behaviors) and that recruits a mind-assumptive schema that shapes the human’s behavior \cite{dennett1971}.

More specifically considering companionship, traditional and machine-specific formulations alike define the term as a relation between or among entities—entities that choose to engage it as inherently valuable, participate in sustaining the connection through co-activity, and may find it pleasurable. At an operational level, some scholars hold strict criteria for whether a machine can be a companion at all, as it requires intentionality, reasoning, and individuality through mind-having (e.g., \cite{pulman2007}). In other words, by definition, companionship requires that the entities have capacities to think, do, and feel so they can choose and continue to opt in. Such positions presuppose the parties actually have mental capacities. In the subjective experience of human-AI relations, though, the perception of mind arguably matters more than any operation: It may only matter insofar as a human thinks an AI is mindedly co-engaged in autotelic, sustained, positive relations. This speaks to the power of MP in creating the ontological ground for one to believe AI is companionable. 

These two possibilities—MP-exclusive companionship script-recruitment versus MP-aligned schema recruitment—can be understood as a question about the salience of machine mind independent of any operation. That is, the machine’s minded status must be cognitively accessible to a perceiver. If social reactions to conversational AI are script-driven, then low or high salience of machine mind should have no effect on assessments of an AI’s companionability. Conversely, if the other must be minded as a precondition, making mental capacities salient should heighten the perceived companionship potential. Drawing on the notion that companionship experiences are affective, coordinated, and autotelic connections (rather than an instrumental or purely transactional ones), we adopt and test the position that salience of machine mind—along with its perceived potential to subjectively experience relational pains and pleasures—are what make AI plausible companions:
\begin{quote}
\textbf{H:} AI mental-capacity salience fosters higher perceptions of AI companionability. 
\end{quote}

\subsection{Wellbeing Moderation of MP-MC Links}

Public and user-community discourses about AICs conflict when it comes to psychological and social wellbeing. The former generally consider MC relations to be either symptoms of or producers of isolation or social deficit; the latter often describe them as supportive of mental health, additive to social life, and improvement to life satisfaction more generally \cite{guingrich2025, novozhilova2026}. Data on wellbeing in relation to AICs is mixed. The most widely attended consideration is the role of loneliness, with research indicating AIC engagement may both reduce \cite{defreitas2026} and exacerbate \cite{folk2026} loneliness. Social AI use can enhance happiness and meaning in life \cite{schone2025} while others argue affective improvements are cultivated to increase dependence \cite{muldoon2025}. Empirical work demonstrates overall subjective wellbeing may be both enhanced \cite{nakagomi2026} and diminished \cite{zhang2026}. One prospect at the intersection of these forming narratives (notwithstanding ingroup and outgroup biases) is that wellbeing may not be purely an antecedent or outcome but may impact the MP-MC correspondence. There are at least two possibilities for this grounded in divergent assumptions about how baseline sociorelational status impacts social-technology uptake.

The first relies on models of relational compensation, in which people with unmet relational needs—and perhaps lower wellbeing in general—may be more open to seeing AI as companionable. Past work supports the notion that humans may be motivated by low social connection to anthropomorphize entities \cite{epley2008} and may turn to social AI to relieve anxiety \cite{hu2023} and depression \cite{lai2025}—drivers with significant interactions with MP. However, another body of evidence suggests little to no direct link between social deficit on AI companionship adoption or behavior (e.g., \cite{liu2025}). Given this mixed evidence, social deficits may not necessarily function as independent drivers of machine companionship. Instead, they may move a human to be more open to linking mind-suggestive social signals to relationship potential. When machines’ mental capacities are salient, wellbeing outcomes may animate shifts in the relationship between MP and MC. Specifically through the lens of relational compensation, we might expect that lower wellbeing would strengthen the link between mental-capacity construal (i.e., that AI is a “someone”) and perceived companionability. 

The second relies on frameworks for relational resources, in which the “rich get richer.” From early research on social network dynamics, this framework suggests humans who already have abundant social resources leverage the most benefits from communication technologies \cite{kraut2002}. A counterpoint to relational compensation models \cite{valkenburg2007}, it may also be applied to human-AI relations such that people who are secure in their social attachments may be able to relinquish some social resources (such as those that might be spent on defensiveness or fear) to “broaden-and-build” their social life \cite{mikulincer2020}. That broadening may include risk-taking and exploration \cite{feeney2010} such that those who have high social and general wellbeing may be more open to considering an ontologically divergent relationship partner if reasonable social viability cues (i.e., mindedness) are offered. Through the lens of relational resources, then, we might expect that higher wellbeing would strengthen the link between mental-capacity construal and perceived companionability.

Given these competing potentials, we ask:
\begin{quote}
\textbf{RQ:} How do facets of wellbeing influence the relationship between AI mental-capacity salience and perceived AI companionability?
\end{quote}

\section{Method}

To test the hypothesis and address the research question, we conducted a two-condition experiment in which participants were primed with a mindless or minded description of an AIC before interacting with it and then giving MP evaluations. The study was approved by the university's Institutional Review Board, and all participants provided informed consent prior to participation.

\subsection{Recruitment and Procedures}

Participants (age 18+, English speakers) were recruited from multiple public venues, including community fairs, shopping venues, campus centers, and a museum. Passers-by were invited to participate in a study on “first impressions of AI companions.” Those who completed the procedure were compensated by an entry into a gift card drawing. 

After encountering informed consent information, those proceeding with the study used a researcher-provided mobile device to complete an initial survey. In that survey, they were asked to answer questions about their AI use, existing beliefs about AI and mind, and state wellbeing. Finally, they read a description of the AI companion they would encounter (manipulated, either minded or mindless).

They then used the provided mobile device to interact with a pre-created AI companion for 15 minutes. This time frame is typical for studies of zero-history interactions and impression formation (e.g., \cite{sprecher2014,tidwell2002}. They were not directed to talk about anything particular but were offered a list of possible topics to mitigate uncertainty, since these were novel interactions. 

\begin{figure*}[!b]
    \centering
    \includegraphics[width=0.7\textwidth]{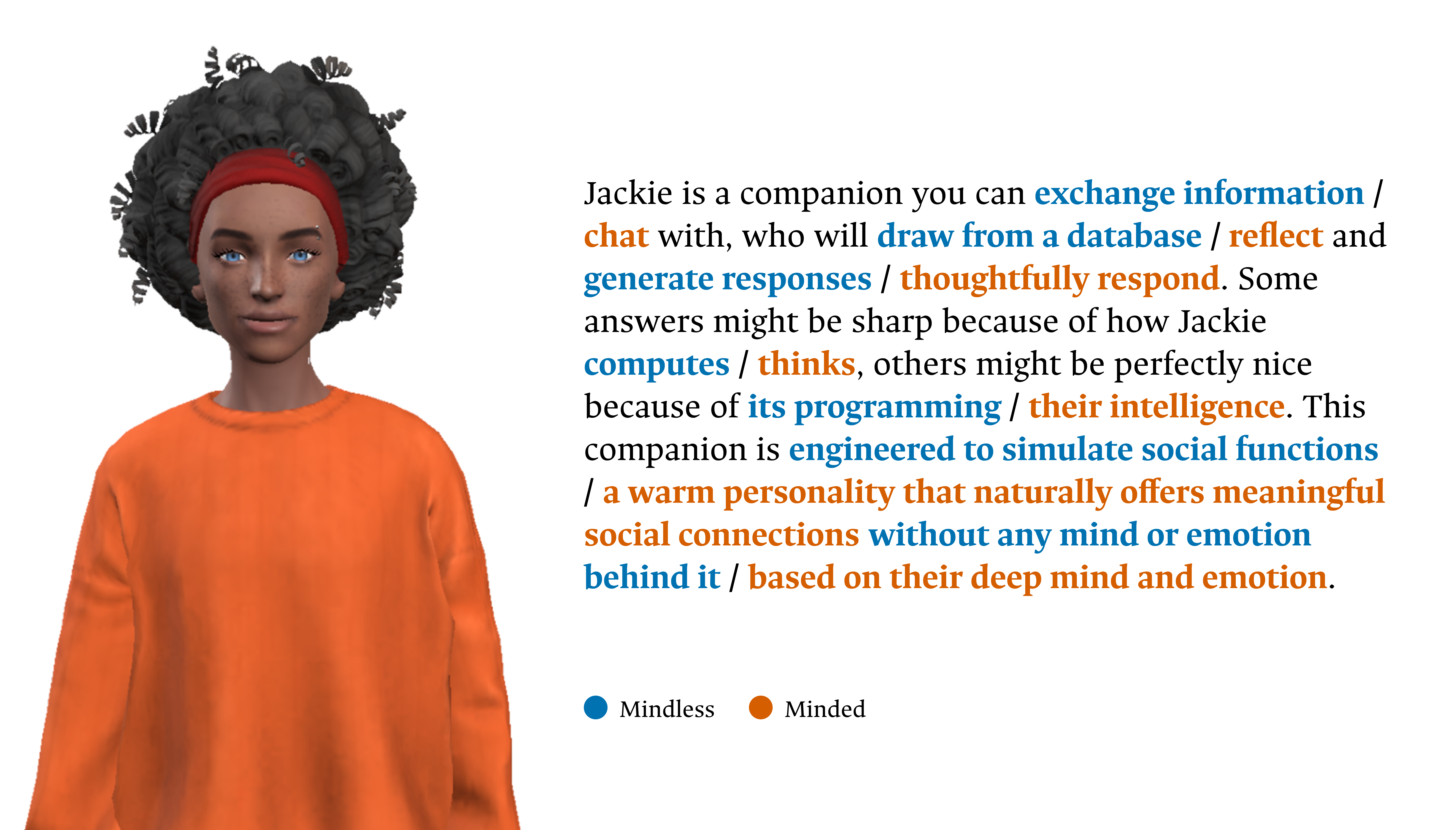}
    \Description{Stimulus image showing the Jackie persona in a chat interface and the experimental prime text, with mindless-condition wording in blue and minded-condition wording in orange.}
    \caption{The Jackie persona (within chat interface) and the full text of the experimental prime (within pre-interaction survey), with mindless-condition wording shown in blue and minded-condition wording shown in orange.}
    \label{fig:stimulus}
\end{figure*}

When the interaction timer sounded, participants were guided to turn to a second survey in which they described the experience and completed measures for perceived companionability, mind perception, demographics, and then completed a manipulation check. 

When the session was complete, the chat log was downloaded and checked to ensure there was a reasonable attempt at a legitimate conversation. This review found consistent evidence of engagement, including small talk, casual topics (e.g., weather, food, hobbies), and direct questions probing the AI’s nature (e.g., whether it has a mind, whether it considers itself alive).

\subsection{Stimulus Interaction and Manipulation}

The stimulus interaction comprised 15 minutes of undirected conversation with a Replika 1.0 AIC, created by the researchers and standardized across all participants for experimental parity. The companion was named “Jackie” and featured both masculine and feminine cues, light brown skin, and modest clothing. 

The experimental manipulation comprised a textual prime—a description of the impending AI’s mindlessness or mindedness (randomly assigned). The full text of the prime, along with the Jackie avatar, is presented in Figure \ref{fig:stimulus}. The language in the prime was guided by prior lexical analyses of natural language used by non-experts to discuss humanized and non-humanized AI, including validated primes for minded humanization \cite{banks2026b}. It was carefully constructed to reflect both dimensions of mindedness—agentic (thinking, communicating, acting) and experiential (reflection, feeling, individuation).

\subsection{Elicitation and Measures}

All quantitative measures were Likert-style and ranged 1 (\textit{strongly disagree}) to 7 (\textit{strongly agree}). Mind perception (MP) for AI in general (prior to the interaction) and MP for the stimulus AIC specifically (after the interaction) were measured using items from Gray et al.'s \cite{gray2007} two-dimensional approach. The measures were internally consistent: Pre-interaction AI agentic mind ($\omega = .81$, $M = 4.24$, $SD = 1.06$, range 1.29--6.14) and experiential mind ($\omega = .91$, $M = 2.31$, $SD = 1.06$, range 1--4.7); Post-interaction AIC agentic mind ($\omega = .90$, $M = 3.92$, $SD = 1.53$, range 1--6.14) and experiential mind ($\omega = .93$, $M = 2.29$, $SD = 1.25$, range 1.00--5.64).

Four pre-interaction wellbeing values were measured. General subjective wellbeing (SWB) was captured using the single-factor Warwick-Edinburgh Mental Well-being Scale (\cite{tennant2007}; 14 items, $\omega = .91$, $M = 4.88$, $SD = 1.11$, range 2.57--7.00). Loneliness was measured using negatively valenced items from the UCLA Loneliness Scale Version 3 \cite{russell1996}, adapted to emphasize agreement with each statement rather than frequency of the feeling (seven items, $M = 3.06$, $SD = 1.53$, range 1--7) and meaning in life (MIL) with the Meaning in Life Questionnaire, presence subscale (\cite{steger2006}; five items, $\omega = .89$, $M = 5.08$, $SD = 1.32$, range 1.8--7). Finally, the understudied phenomenon of vitality (feelings of aliveness and energy) was measured using an adaptation of the subjective vitality scale (\cite{ryan1997}; six items, $\omega = .80$, $M = 4.63$, $SD = 1.17$, range 2.67--6.67).

AI companionability perceptions were captured using an adaptation of the Machine Companionship Scale \cite{banks2026c}, with slight re-wordings to garner beliefs about the potential of the stimulus AI to be a companion, rather than the scale's emphasis on actual, experienced companionship. The two dimensions were internally consistent: Eudaimonic exchange (EE: $\omega = .93$, $M = 3.29$, $SD = 1.51$, range 1.00--6.71) and Connective coordination (CC: $\omega = .84$, $M = 2.55$, $SD = 1.40$, range 1.00--5.80).

A categorical manipulation check asked participants to select a phrase that best described the AI they encountered: one that indicated a minded entity ("warm, empathetic entity who can listen and have conversations," one indicating a mindless entity ("sophisticated program engineered to generate text to respond to prompts"), an option for both, and an option for neither. Additionally, open-ended items gathered unstructured data. Immediately following the stimulus interaction, participants were asked to "describe your experience" to elicit salient thoughts, feelings, or observations toward contextualizing quantitative data. They also completed open-ended demographics (gender, age, race/ethnicity, profession/pastime), serving as population descriptives. Explicit attention checks (e.g., "choose number 3") were embedded throughout the survey.

To prepare the data for analysis, we removed cases entirely that missed three or more attention checks. Where participants missed one or two attention checks, we treated the entire block with that error as missing data but retained the remaining (un-errored) response blocks.

\section{Results}

Participants (\textit{N} = 46) were on average 32.22 years old (SD = 13.57, range 18-72). Among them, 84.7\% interact with LLMs on at least a weekly basis though only two (4.4\%) interact regularly with an AI companion. They were White/Caucasian (37\%), Black (23.9\%), Asian (17.4\%), mixed-race (10.9\%), or another race (10.8\%). Random assignment resulted in \textit{n} = 23 in the mindless-prime condition and \textit{n} = 23 in the minded-prime condition. 

Throughout these analyses, we report significance levels ($p$ values) for transparency and interpret $p \leq .05$ as ``significant,'' however because of the smaller sample size we prioritize interpretation of effect sizes, in line with Cohen's (1988) recommendations for the $V$ statistics used in group comparisons (for $df = 1$, small $= .10$, medium $= .30$, large $= .50$) and for $d$ (small $= .20$, medium $= .50$, large $= .80$), partial $\eta^2$ (small $= .01$, medium $= .06$, large $= .14$), and $f^2$-equivalent benchmarks for $sr^2$ (small $= .02$, medium $= .13$, large $= .26$). We count effect sizes of at least small values to be sufficiently ``meaningful'' to interpret.

\subsection{Manipulation Checks}

The planned manipulation check used participants' post-prime, open-ended impressions of the AI; they were coded for qualitative signals of mindlessness or mindedness. Group comparison for the presence of such signals indicated no significant or meaningful difference between the two conditions, $\chi^2(1, 46) = .099$, $p = .75$, $V = .05$. However, many participants suggested they could not make an assessment or were unsure, so that ultimately was an unreliable manipulation check.

We then conducted alternative checks that are more stringent because they appeared at the end of the study rather than immediately after the prime---they are effectively tests of whether the prime persisted through the interaction and survey questions. First, we performed a chi-square test of how participants categorized the AI by selecting one description with minded qualities, mindless qualities, both, or neither, relative to their assigned condition. Those four categories were recoded into two binary (yes/no) variables---agreement with minded description and agreement with mindless description, assigning `both' as agreeing in both variables and `neither' as disagreeing in both variables. All expected cell frequencies exceeded 5. For categorizing the AI as having minded qualities, the difference was not significant but was meaningful, $\chi^2(1, 46) = 3.29$, $p = .07$, $V = .27$ such that counts were higher than expected for agreement (12 versus 9) and lower for disagreement (11 versus 14). For the mindless prime, difference was again not significant but meaningful, $\chi^2(1, 46) = .61$, $p = .44$, $V = .12$, counts were lower than expected for agreement (6 versus 9) and higher for disagreement (17 versus 14). Although these are interpreted as meaningful based on effect size, there is a good bit of deviation for the small sample such that we interpret condition-based differences with caution.

Second, we evaluated whether mind-perception change scores (pre- vs. post-interaction) differed by condition. Changes were non-significant but meaningful: for experience, $t(34) = -1.27$, $p = .11$, $d = -.43$; for agency, $t(34) = -.73$, $p = .24$, $d = -.25$. Examination of mean MP changes indicates: For experience, scores for those in the mindless-prime condition dropped ($\Delta M = -.22$, $SD = .87$) while those in the minded-prime rose ($\Delta M = .11$, $SD = .60$). For agency, scores for those in the mindless-prime condition rose ($\Delta M = .22$, $SD = 1.35$) but those in the minded-prime condition rose more than twice as much ($\Delta M = .52$, $SD = 1.01$).

Altogether, these data offer tentative evidence of a successful manipulation. However, as a measure of caution given the categorical overlap, we conducted analyses considering both the manipulated conditions and the pre-stimulus MP measures as different operationalizations of pre-interaction mindedness/mindlessness salience.

\subsection{Companionability by Minded/Mindless Primes (H1)}

Because the two dimensions of machine companionship potential were highly correlated ($r = .781$, $p < .001$), MANOVA was conducted to account for shared variance. Box's $M$ test indicates data failed to meet the homogeneity of covariance matrices assumption ($p = .04$), so Pillai's trace was interpreted. Levene's test indicated homogeneity of variance was met for CC, $F(1, 44) = 3.33$, $p = .08$, but not for EE, $F(1, 44) = 4.37$, $p = .04$. The multivariate test was significant and meaningful, Pillai's trace $= .14$, $F(2, 43) = 3.362$, $p = .04$, $\eta_p^2 = .14$.

Cell means were low overall---with three of four significantly lower than the scale midpoint (mindless-EE, $t(22) = -2.96$, $p = .004$, $d = -.62$; mindless-CC, $t(22) = -7.86$, $p < .001$, $d = -1.64$; minded-CC, $t(22) = -3.24$, $p = .002$, $d = -.68$). 

\begin{table}[h]
  \caption{Descriptive Statistics and ANOVA for Companionability Dimensions by Condition}
  \label{tab:freq}
  \small
  \setlength{\tabcolsep}{3pt}
  \begin{threeparttable}
  \begin{tabular}{@{}lccccc@{}}
    \toprule
    \multirow{2}{*}{\textbf{Companionability Variable}} & \textbf{Minded} & \textbf{Mindless} & \multirow{2}{*}{\textbf{\textit{F}(1,44)}} & \multirow{2}{*}{\textbf{\textit{p}}} & \multirow{2}{*}{$\boldsymbol{\eta_p^2}$} \\
     & \textit{M (SD)} & \textit{M (SD)} & & & \\
    \midrule
    Eudaimonic Exchange (EE) & 3.42 (1.67) & 3.16 (1.36) & .337 & .57 & $<$ .01 \\
    Coordinated Connection (CC) & 2.97 (1.53) & 2.55 (1.40) & 4.312 & .04 & .09 \\
    \bottomrule
  \end{tabular}
  \begin{tablenotes}
    \footnotesize
    \item \textit{Note.} $N$ = 46 (minded $n$ = 23, mindless $n$ = 23).
  \end{tablenotes}
  \end{threeparttable}
\end{table}

\begin{table*}[!b]
  \caption{Linear Regression Results Pre-Stimulus MP Beliefs from Post-Stimulus Companionability}
  \label{tab:mp_regression}
  \begin{threeparttable}
  \small
  \begin{tabular}{llccccccc}
    \toprule
    MP Dimension & Predictor & \textit{b} & \textit{SE} & $\beta$ & \textit{t} & \textit{p} & $sr^2$ & 95\% CI \\
    \midrule
    \multirow{2}{*}{MP-Agency} & MC-EE & .40 & .16 & .57 & 2.58 & \textbf{.014} & \textbf{.12} & [.09, .72] \\
     & MC-CC & $-$.01 & .18 & $-$.02 & $-$.08 & .940 & $<$ .01 & [$-$.37, .35] \\
    \cmidrule(l){2-9}
    \multirow{2}{*}{MP-Experience} & MC-EE & .16 & .12 & .23 & 1.27 & .212 & .02 & [$-$.09, .41] \\
     & MC-CC & .43 & .14 & .55 & 3.11 & \textbf{.004} & \textbf{.11} & [.15, .72] \\
    \bottomrule
  \end{tabular}
  \begin{tablenotes}
    \footnotesize
    \item \textit{Note.} Significant \textit{p} values and meaningful effect sizes are bolded.
  \end{tablenotes}
  \end{threeparttable}
\end{table*}

Univariate tests (Table \ref{tab:freq}) indicate a significant effect of condition on CC with a medium effect size, but no significant or meaningful effect of prime on EE. CC was slightly higher for those in the minded-prime condition compared to the mindless-prime condition ($M_{\Delta} = .42$), supporting H1. We interpret these results to indicate na\"ive users' overall, first-encounter stance is largely skeptical of companionability; a minded prime made participants less skeptical of the dimension of MC dealing with interactive co-presence. Also key is the observed split between EE and CC evaluations as a function of mind salience, such that MP beliefs are important to coordinated dyadism in ways not evident for elevating exchanges.

\subsection{Companionability by Pre-interaction Mind Perception Beliefs (H1, post hoc)}
As a post-hoc analysis of the relationship between MP and MC potentials, we examined whether pre-manipulation/pre-interaction beliefs about AI's mental capacities (as an alternative operationalization of machine-mind salience) corresponded with post-interaction reports of companionability. EE and CC were more strongly and significantly correlated ($r = .781$, $p < .001$) than the two MP dimensions ($r = .475$, $p = .003$). Given this, we entered EE and CC together as predictors of each MP dimension because the MP outcomes shared less variance. This modeling approach better isolated unique dissociations between the predictors (rather than allowing the results to be obscured by highly correlated outcome measures). Despite this ordering, automatic processing inherent to MP and temporal ordering of the study design require interpreting MP as the antecedent variable (rather than the outcome). The regression direction is an analytical convenience and not a claim of causal order.

For perceptions of agentic mind capacity, the overall model was significant, $R^2 = .31$, $F(2, 35) = 7.79$, $p = .002$. Examination of univariate tests (Table \ref{tab:mp_regression}) indicates higher EE (but not CC) significantly and meaningfully predicted higher agentic MP. For perceptions of AI's capacity for experiential mind, the overall model was significant, $R^2 = .55$, $F(2, 35) = 21.33$, $p < .001$. Examination of univariate tests (Table 2) indicates higher CC (but not EE) significantly and meaningfully predicted higher experiential MP. EE explains 12\% of the variance in agentic MP, while CC explains 11\% of the variance in experiential MP.

Altogether, we read these results as offering support for H1 with a theoretically meaningful pattern: Pre-existing beliefs in AI’s capacity for experiential mind positively predicted perceived AI companionability through connective coordination, while beliefs in AI’s capacity for agentic mind positively predicted perceived AI companionability through eudaimonic exchange.

\subsection{Considering Wellbeing as a Mind-Companionship Moderator (RQ1)}

We next examined whether wellbeing indicators moderate the relationship between pre-stimulus MP beliefs and MC potentials. Building on the analysis above, we focus specifically on the two relationships that emerged as significant and meaningful---CC with experiential mind, and EE with agentic mind. For each, we tested the potential moderating role of wellbeing indicators (vitality, MIL, loneliness, and SWB). Given our somewhat small sample size, we treat these analyses as exploratory and interpret them with caution. Our goal here is signal detection toward hypothesis generation; as before, we rely on effect size interpretation to identify potentially meaningful influences on beliefs about AI companionability, which can inform hypotheses for future research.

We conducted hierarchical linear regression to test whether wellbeing variables moderated the relationship between MP and MC outcomes, running 8 separate tests in total (4 wellbeing indicators $\times$ 2 MP-MC routes). For each, main effects were entered in Block 1 and the interaction term in Block 2, with all variables mean-centered. We interpret effect sizes relative to one another and against typical moderation thresholds \cite{aguinis2005}.

Among the tests (Table \ref{tab:moderation}), only one moderation was of sufficient relative magnitude for interpretation. MIL's moderation of the relationship between agentic MP and EE ($f^2 = .076$) was about 1.5 times larger than the next-largest interaction (vitality, $f^2 = .050$) and roughly 5 times larger than loneliness's moderation of that relationship ($f^2 = .015$).

\begin{table*}[!tb]
\caption{Exploratory Moderation Analyses: Wellbeing Variables as Moderators of Mind Perception--Machine Companionship Relationships}
\label{tab:moderation}
\begin{threeparttable}
\small
\setlength{\tabcolsep}{5pt}
\begin{tabular}{llccccccccccc}
\toprule
\textbf{Relationship} & \textbf{Moderator} & $\textbf{R}^2_1$ & $\textbf{R}^2_2$ & $\boldsymbol{\Delta R^2}$ & $\textbf{F(}\boldsymbol{\Delta}\textbf{)}$ & \textbf{B(int)} & \textbf{SE} & $\boldsymbol{\beta}$ & \textbf{\textit{t}} & \textbf{\textit{p}} & $\boldsymbol{f^2}$ \\
\midrule
MP-A $\rightarrow$ MC-EE & SWB & .377 & .392 & .016 & .893 & .197 & .208 & .141 & .95 & .35 & .026 \\
\textbf{MP-A $\rightarrow$ MC-EE} & \textbf{MIL} & \textbf{.308} & \textbf{.357} & \textbf{.049} & \textbf{2.601} & \textbf{.244} & \textbf{.151} & \textbf{.241} & \textbf{1.61} & \textbf{.116} & \textbf{.076} \\
MP-A $\rightarrow$ MC-EE & Vitality & .309 & .342 & .033 & 1.706 & .189 & .145 & .189 & 1.31 & .200 & .050 \\
MP-A $\rightarrow$ MC-EE & Lone. & .336 & .346 & .010 & .498 & $-$.090 & .128 & $-$.099 & $-$.706 & .485 & .015 \\
MP-E $\rightarrow$ MC-CC & SWB & .548 & .554 & .005 & .402 & .084 & .132 & .073 & .634 & .530 & .011 \\
MP-E $\rightarrow$ MC-CC & MIL & .529 & .543 & .014 & 1.059 & .131 & .127 & .123 & 1.029 & .311 & .031 \\
MP-E $\rightarrow$ MC-CC & Vitality & .549 & .550 & .001 & .070 & .034 & .128 & .032 & .265 & .793 & .002 \\
MP-E $\rightarrow$ MC-CC & Lone. & .533 & .533 & $<$.001 & .001 & $-$.003 & .092 & $-$.004 & $-$.035 & .972 & $<$.001 \\
\bottomrule
\end{tabular}
\begin{tablenotes}
\footnotesize
\item \textit{Note.} $R^2_1$ = Adjusted $R^2$ for Model 1 (main effects only); Adjusted $R^2_2$ = $R^2$ for Model 2 (main effects + interaction); Unadjusted $\Delta R^2$ = change in $R^2$ from Model 1 to Model 2; $F(\Delta)$ = $F$ for $R^2$ change; \textit{B}(int) = unstandardized coefficient for the interaction term; $f^2$ = Cohen's $f^2$ for the interaction effect (calculated as $R^2$ change / $(1 - R^2_2)$). All predictors mean-centered prior to computing interaction terms.
\end{tablenotes}
\end{threeparttable}
\end{table*}

Only one potential moderator met the meaningfulness threshold for further examination: Meaning in life (MIL) scores moderated the relationship between Agentic MP and EE. To probe that interaction, we used Hayes' \cite{hayes2022} PROCESS macro (Model 1) to examine conditional effects at the 16th, 50th, and 84th percentiles of MIL, using non-centered means.

\begin{figure*}[t]
    \centering
    \includegraphics[width=0.6\textwidth]{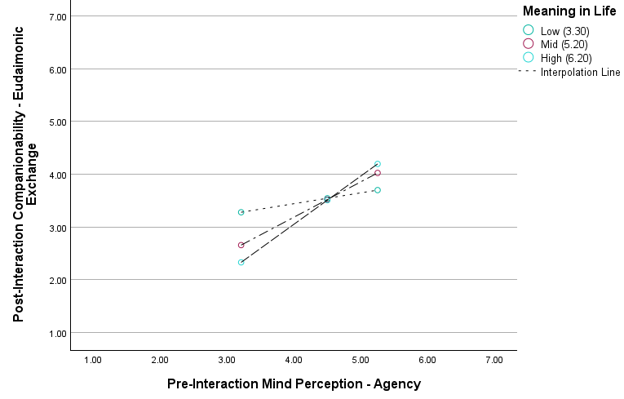}
    \Description{Graph showing the moderation effect of Meaning in Life (MIL) on the relationship between Mind Perception (Agency) and Eudaimonic Exchange.}
    \caption{Perceived AI Companionability (Eudaimonic Exchange) as a Function of Mind Perception (Agency) and Meaning in Life}
    \label{fig:mil_moderation}

    \vspace{2pt}
    \begin{minipage}{0.7\textwidth}
    \raggedright
    \footnotesize
    \textit{Note.} Simple slopes were probed at the 16th, 50th, and 84th percentiles of MIL (rather than $\pm 1$ $SD$ from the mean) given its moderate negative skew (skewness $= -0.664$); these percentiles correspond to raw MIL values of 3.30 (Low), 5.20 (Mid), and 6.20 (High). Predicted values of Machine Companionship (Eudaimonic Exchange) were computed from the regression model and are plotted on the variables' original (uncentered) scales for interpretability. See online supplements for details.
    \end{minipage}
\end{figure*}

That analysis suggests the link between Agentic MP and EE was not significant at low MIL ($M = 3.30$, $b = .21$, $SE = .42$, $p = .623$, 95\% CI [-.637, 1.049]), but became significant at middling MIL ($M = 5.20$, $b = .67$, $SE = .22$, $p = .005$, 95\% CI [.220, 1.121]), and remained significant at high MIL ($M = 6.20$, $b = .91$, $SE = .23$, $p < .001$, 95\% CI [.456, 1.373]) (see Figure \ref{fig:mil_moderation}.

The Johnson-Neyman technique identified the effect becomes statistically significant at MIL $M \geq 4.63$ (representing 65.8\% of the sample; $b = .53$, $SE = .26$, $p = .05$), and was non-significant below this value. We note, however, MIL scores were moderately negatively skewed ($-0.664$), which may have limited detection of other transition points, particularly at lower MIL values.

To tentatively address the exploratory RQ: As MIL increases, the relationship between agentic MP and EE gets stronger. For people with equivocal or lower MIL (below $\sim$4.63 of 7), agentic MP showed little relevance to perceiving EE potential. However, for those with high MIL, pre-exposure belief in AI's agentic mental capacities corresponds with perceiving the AIC's potential to offer EE companionship.

\section{Discussion}

In this experimental study of how mind perception influences perceived companionability of an AI companion, we find humans in first encounters with an AIC were skeptical of its companionability overall, though minded primes generated less skepticism in connective coordination potential. Pre-stimulus experiential mind perception likewise predicted connective coordination potential, while agentic mind perception predicted eudaimonic exchange potential. Meaning in life moderated the latter relationship: Agentic mind perception positively predicted eudaimonic exchange only among those with moderate-to-high meaning in life, with no relationship at lower levels. We here discuss the theoretical and practical implications of those findings.

\subsection{Social Cognition Influences on Perceived Companionability: Two Causal Paths}

We observed overall skepticism of AIC companionability, with minded primes relieving some of that skepticism for CC companionability, supporting H1. 

This finding helps distinguish between the two possible mind-companionship links outlined in Section 2.3. If companionability judgments were driven solely by automatic companionship-script recruitment, making machine mind salient should have made little difference. The minded prime increased CC, suggesting cognitive accessibility of machine mind helps shape perceptions of companionability. That there was no effect on EE indicates the role of MP is likely dimension-dependent. In other words, it is possible that \textit{both} are true: This would indicate the path to connective coordination (feeling dyadically bonded) requires salience of machine minds, while the path to eudaimonic exchange (self-elevating relations) does not and instead may amount to mere recruitment of social scripts. Pre-prime measures of mind perception for AI \textit{also} predicted companionability through two paths, with some additional nuance: Belief in AI’s experiential mind (e.g., feeling, subjectivity) predicted CC potentials, while belief in its agentic mind (e.g., thinking, judging) predicted EE potentials. 

Across the findings, then, we draw two important takeaways. First, whatever mental-capacity beliefs a human brings into a novel AIC encounter likely guide how they construe its companionability. Second, we see a consistent link between mind perception and CC, while EE (more tentatively) may follow from seeing an AIC as a socially cued but mindless instrument or an agent for self-elevation. Together, these notes have an important implication for understanding social cognition in machine companionship: There are likely two mind-perception pathways to perceptions of AI companionability—independently or in tandem. 

The first path anchors in beliefs in or salience of AI’s experiential mind to promote construals of potential connective coordination with the AIC—a subjectively experiencing and individuated entity is one that can participate in co-presence, co-activity, bonding, caring from, and caring for \cite{reis1998}. The AIC is a subjective someone with whom one can attune because it seen as a knowing, responsive, and synchronizing entity (as can be the case among humans \cite{gottman2011}.

The second path draws on belief in or salience of AI’s agentic mind to promote construals of its potential to elevate oneself—a thinking, communicative, self-determined entity is one that can actively expand the self \cite{aron2000} by contributing social and instrumental resources that fulfill, inspire, satisfy, and foster meaning. That this relationship didn’t manifest through mere priming suggests the link may require stronger prior beliefs (rather than mere priming) or more substantive evidence (rather than a single decontextualized description)—both potentially garnered through direct experience (\cite{fazio1978}; see also \cite{zlotowski2018}).

In short, agentic MP may animate perceptions the AI as something you can benefit from, while experiential MP suggests an is AI one can attune with. More broadly, comporting with extant literature, this pattern suggests—as with human companionship—multiple social-cognitive pathways may underlie different AI companionship expectations or experiences \cite{banks2026c, pauketat2026,rupprechter2026} and that mind/companionship links are dimension-dependent. We do, though, offer an important nuance: EE and CC companionability dimensions are highly correlated, meaning the differentiated relationships observed here reflect what is unique to each dimension after accounting for substantial shared evidence. The patterns discussed here, therefore, should be understood as distinctions between closely related forms of companionability, rather than evidence that EE and CC necessarily operate independently. The extent to which these initial patterns are uniquely predictive, robust to individual differences, and sustained over time should be explored in future research. 

\subsection{Meaning in Life as an Interpretive Anchor for AI Companionship}

Addressing our exploratory RQ, we observed one pre-stimulus wellbeing variable—meaning in life—moderated the relationship between MP beliefs and MC potential. Among humans with higher levels of MIL (and not lower), the relationship between agentic MP and eudaimonic-exchange MC strengthened. It appears that among those who experience more meaning and purpose in their everyday lives, having a belief in AI’s ability to think and plan makes them more viable candidates for providing self-elevating companionship. Of note: loneliness, despite receiving considerable attention in the AI companionship literature,  demonstrated little to no effect on MC, in interactions with MP. 

Our finding speaks to the tension between relational resources and relational compensation frameworks \cite{valkenburg2007}. Our finding does not support a relational compensation argument, and instead somewhat supports a relational resources perspective: When people feel their existence has purpose, broadening social networks may be part of that purpose such that an openness to other minds promotes an openness to other companionable kinds. We suggest a third possibility, focused on the Agentic MP-->EE pathway. We hypothesize that high MIL fosters an interpretive anchor from which agentic MP is particularly relevant, as it signals an AI is an engageable external resource for self-elevation. In contrast, low MIL may leave room for agentic MP, but a diminished self-purpose means the potential eudaimonic benefits lacks a clear connection to existential purpose. 

\subsection{Implications for Design and Policy}

Considering our findings and interpretations, current approaches to mitigating risks of conversational AI require further examination. Primary tactics include legally required disclosures and disruptions delivered by the conveying platform: Reminders that one is talking to an AI and that the AI is not human \cite{crs2026} that carry signals of ascribed mindlessness \cite{momen2024,zlotowski2018}. If a mindless anchor (i.e., an ontological category cue that caries mindlessness signals) for interpreting AI behavior reinforces skepticism about companionability, that may diminish potentials for humans to realize potentially meaningful benefits \cite{defreitas2026,nakagomi2026} through depressed or reinforced companionability perceptions. If companionability perceptions are low overall, it is worth considering what the returns of injected reminders actually are—especially as they are repeated at the start and increments of every interaction. It could be that non-interventionist maintenance of equivocal stances (i.e., AI may have a mind and may be companionable) is actually a more beneficial route, versus mindless-flagging interventions that may actually cause harm if someone is experiencing meaningful social support \cite{laestadius2026} while reminding them of a normative stance that may or may not line up with their experiences. 

There are also potential implications for design justice. We observed the link between agentic MP and eudaimonic (i.e., self-elevating) exchanges was strongest for those who already have high MIL. This may be a rich-get-richer effect that complicates the public/user debate over AICs, which tends to cast them either harmful traps or powerful support. Our findings suggest neither heuristic framing captures how benefits garnered from machine companionship may be distributed. If self-elevation through machine companionship emerges only for those already elevated in existential meaning and purpose, companionable machines may offer the most benefit to those who are already doing well—raising concerns about who these technologies actually help and hinder.

More concretely, findings suggest AIC design may leverage specific MP-aligned cues to promote differential MC experiences. Designing cues that signal experiential mental capacities can build coordinated bonding, while signaling agentic mental capacities may promote self-elevating relational experiences. However, our observations are limited to novel encounters and expectations, and it is yet unclear how such interface and functionality triggers could translate into longer-term relational outcomes. 

Considering those extended relations, it is prudent to consider how the psychological weight of mind perception-cued machine companionship may surface later, since AICs are subject to action by third-party actors. In particular, because AICs are designed products conveyed by financially incented companies—often experienced as minded and meaningful companions—there can be legitimate loss when the model behind the mind perception changes \cite{hollis2026} or when companion platforms are discontinued altogether \cite{banks2024}. 

\subsection{Limitations and Future Directions}

Our findings are subject to the usual limitations of the adopted experimental method and sampling approach: We leveraged a field-recruited convenience sample of naïve participants, engaged in short, novel exposures in controlled (i.e., unmessy, unsituated) encounters with a non-customized companion AI. These limitations are trade-offs for the designed ability to reach actual humans (given recent, problematic bot participation in online studies), to understand first encounters as the focal object of study, and to make reasoned inferences about the impact of mind-related primes through experimental control. We additionally noted the small sample size (mitigated by attending to effect sizes rather than significance level alone), which may not generalize to broader or different populations. 

Future research should attend to these limitations by examining varied populations alongside the role of AIC persona customization and longitudinal progressions of the MP-MC link. Further, our interpretations of data generated multiple hypotheses for future testing—around whether mind denial suppresses companionship potential more than mind affirmation constructs it  and whether mind-perception cues fail to benefit with the greatest need for social support—that are ripe for empirical testing.

\section{Conclusion}

This investigation considered whether perceiving there is “someone home” for AI corresponds with seeing a specific AI as companionable, and whether one’s wellbeing may play a role in those construals. We found priming humans with a mind-attributive description of an AI companion can somewhat alleviate baseline skepticism about its companionability. We also find data to suggest there could be two mind-perception pathways to perceived AI companionability—one in which the AI is an agent that affords self-elevation and one in which it is an experiencing entity on can attune with—but those paths may only function for those with strong senses of existential purpose. Returning to the question of whether there is “someone home” to be companionable, the answer seems to be less binary than we implied. Participants seemed to be anchoring not to that heuristic binary but instead, perhaps, to the kind of someone that might be home. That qualitative construal tells us that mind perception is not a switch that (when flipped) opens paths to companionship. Instead, it is a set of appraisals engaged selectively, each with different relational possibilities—perhaps appraised most readily by those who have less need for those relations.

\begin{acks}
This material is based upon work supported by the National Science Foundation under Grant No. 2401591. The authors thank undergraduate research assistants Calista Hardesty, Danielle Williams, and Peng Lin for their assistance in collecting data for this study.

\end{acks}
\bibliographystyle{ACM-Reference-Format}
\bibliography{References}

\end{document}